\documentclass[pra,twocolumn, showpacs, preprintnumbers,nofootinbib,amsmath,amssymb, superscriptaddress]{revtex4-1}

\usepackage{graphicx, color, graphpap}
\usepackage{amsmath}
\usepackage{enumitem}
\usepackage{amssymb}
\usepackage{amsthm}
\usepackage{pstricks}
\usepackage{float}
\usepackage{multirow}
\usepackage{hyperref}

\long\def\ca#1\cb{} 

\newcommand{\tr}{{\mathsf{tr}}}

\renewcommand{\geq}{\geqslant}

\newcommand{\be}{\begin{equation}}
\newcommand{\ee}{\end{equation}}
\newcommand{\ba}{\begin{array}}
\newcommand{\ea}{\end{array}}

\newtheoremstyle{example}{\topsep}{\topsep}%
{}
{}
{\bfseries}
{.}
{   }
{\thmname{#1}\thmnumber{ #2}}
\theoremstyle{example}

\theoremstyle{definition}

\begin{document}

\title{Conservation law for Uncertainty relations and quantum correlations}
\author{Zhihao Ma}
\affiliation{Department of Mathematics, Shanghai Jiaotong
University, Shanghai, 200240, China}
\affiliation{ Department of Physics and Astronomy, University College
London, Gower St., WC1E 6BT London, United Kingdom}

\author{Shengjun Wu}
\affiliation{Kuang Yaming Honors School, Nanjing Univeresity, Nanjing, Jiangsu 210093, China}
\affiliation{Hefei National Laboratory for Physical Sciences at Microscale and
Department of Modern Physics, University of Science and Technology of China, Hefei, Anhui 230026, China}

\author{Zhihua Chen}
\affiliation{Department of Science, Zhijiang college, Zhejiang
University of technology, Hangzhou, 310024, China}
\affiliation{Centre for Quantum Technologies, National University of
Singapore, 3 Science Drive 2, 117543 Singapore}

\begin{abstract}
Uncertainty principle, a  fundamental principle in quantum physics, has been studied intensively via various uncertainty inequalities. Here we derive an uncertainty equality in terms of linear entropy, and show that the sum of uncertainty in complementary local bases is equal to a fixed quantity. We also introduce a measure of correlation in a bipartite state, and show that the sum of correlations revealed in a full set of complementary bases is equal to the total correlation in the bipartite state.  The surprising simple equality relations we obtain imply that the study on uncertainty principle and correlations can rely on the use of linear entropy, a simple quantity that is very convenient for calculation.
\end{abstract}

\maketitle

Through the introduction of quantum states, quantum physics unifies the information of complementary properties in a unique way, and at the same time gives rise to a fundamental phenomenon of quantum physics: uncertainty principle exists for complementary properties.
The more precisely an observable is determined, the less precisely a complementary observable can be known.
The uncertainty principle can be formulated via the standard deviation of a pair of complementary observables, first in the form of the Heisenberg's uncertainty principle $\Delta x \Delta p \geq \hbar /2$ for an infinite dimensional Hilbert space, and later in the form of the Robertson-Schr{\"o}dinger uncertainty inequality for a finite dimensional Hilbert space.

Instead of the standard deviation of observables, the uncertainty principle can also be elegantly formulated in terms of entropies related to measurements in complementary bases. Two bases are unbiased if a basis state in one basis is equally overlapped with all basis states in the other basis. A set of $K$ bases are called mutually unbiased if any pair of them are unbiased. In a $d$-dimensional Hilbert space, at most $d+1$ mutually unbiased bases (MUBs) exist; and when $d$ is a power of a prime number, we know that  $d+1$ MUBs exist for sure. However, it is not clear how many MUBs exists when the dimension $d$ is an arbitrary number, for example, when $d=6$.  When the system is projected onto a certain basis $\{ \left| i_{\theta} \right\rangle | i=1,2,\cdots, d \}$, the uncertainty of the measurement results is nicely characterized by the Shannon entropy $H_{\theta}=\sum_{i=1}^d  -p_i \log _2 p_i$, where $p_i$ is the probability to get the $i$-th basis state $\left| i_{\theta} \right\rangle$. The larger the Shannon entropy $H_{\theta}$, the more uncertain the measurement results. In terms of the Shannon entropies of the measurement results in two MUBs, the uncertainty principle can be formulated as $H_{\theta} +H_{\tau} \geq \log_2 d$ when the two bases $\{\left|i_{\theta} \right\rangle\}$ and $\{\left| i_{\tau} \right\rangle\}$ are mutually unbiased. Thus, the more certain the measurement results in one basis, the less certain the results in its mutually unbiased basis (MUB).  This relation has been further generalized, either to $M$ MUBs, or for other considerations.

An uncertainty relation also exists for a bipartite state $\rho_{AB}$. Suppose $\{\left|i_{\theta}\right\rangle\}$ and $\{\left| i_{\tau} \right\rangle\}$ are two MUBs, and Alice can perform measurement in either of them, then
\begin{equation}
S (\theta | B) + S (\tau | B ) \geq \log_2 d + S(A|B)
\end{equation}
where $S(A|B)$ and $S(\theta | B)$ ($S (\tau | B )$) denote, respectively, the conditional von Neumann entropies of the initial bipartite state $\rho_{AB}$ and the final bipartite state $\rho_{\theta B}$ ($\rho_{\tau B}$) after the measurement in the basis $\{\left|i_{\theta} \right\rangle\}$  ($\{\left| i_{\tau} \right\rangle\}$).

Instead of the Shannon entropy or the von Neumann entropy, we shall use linear entropy to derive uncertainty relations. We shall obtain uncertainty inequalities for any number of MUBs, and surprisingly, also find an uncertainty equality for a full set of MUBs. We shall also introduce a measure of correlation in a bipartite state and present a conservation law of correlations.

The linear entropy of a $d $-level quantum state $\rho$ is defined as
\begin{equation}
 S_{L}(\rho) :=
d(1-\tr(\rho^{2})) . \label{linearentropy}
\end{equation}
Thus, the linear entropy ranges from $0$ (when $\rho$ is a pure state) to $d-1$ (when $\rho$ is maximally mixed).

For a bipartite state $\rho_{AB}$ in a $d \otimes d$ composite Hilbert space, if system A is projected on to the basis $\{\left|i_{\theta} \right\rangle | i=1,2,\cdots,d \}$,
the overall state of the composite system after the measurement on A is given as
\begin{equation}
\rho_{\theta B} =
\sum_{i=1}^d \left| i_{\theta} \right\rangle _A \left\langle i_{\theta} \right|
\otimes _A\left\langle i_{\theta} \right|  \rho_{AB} \left| i_{\theta} \right\rangle _A  . \label{totalstateafterlocalmeasontheta}
\end{equation}
We can introduce the conditional linear entropy
\begin{equation}
S_L ( \theta |B) := S_L(\rho_{\theta B}) - S_L (\rho_B) =d\tr (\rho_B^2) -d^{2}\tr (\rho_{\theta B}^2)+(d^{2}-d)\label{conditionalentropy}
\end{equation}
as a measure of the uncertainty about Alice's measurement result given Bob's state.

When Alice tries to find a basis to perform the measurement on her system so that Bob will know her result with minimum uncertainty, then when she uses another MUB to perform the measurement, Bob will have a large uncertainty about Alice's result. This uncertainty relation is formulated in the following theorem.

{\bf Theorem 1.} For any density matrix $\rho_{AB}$ on a composite Hilbert space $H_{A}\otimes H_{B}$ of dimension $d\otimes d$,
we have the following uncertainty equality
\begin{equation}
\sum_{\theta=1}^{d+1} S_L ( \theta |B)  =   S_L ( A |B)+(d^{3}-d^{2})   \label{UNCERTeq}
\end{equation}
when a full set of $d+1$ MUBs exists for the $d$-dimensional Hilbert space $H_A$. In general, for any number $M$ of MUBs that exist for a $d$-dimensional Hilbert space $H_A$, we always have the following uncertainty inequality
\begin{equation}
\sum_{\theta=1}^{M} S_L ( \theta |B)  \geq  (dM-d^{2})\tr (\rho_b^2) -d (M-1)\tr(\rho_{ab}^2)+M(d^{2}-d) .    \label{UNCERTineq}
\end{equation}

The proof of theorem 1 is left to the Supplementary Information. The fact that an uncertainty equality (\ref{UNCERTeq}) can be obtained for a full set of MUBs implies that linear entropy is the optimum one in investigating uncertainty principle in terms of generalized entropies.

In the following, we shall discuss correlations in bipartite states and present a conservation law of correlations.

For a bipartite state $\rho_{AB}$ with the marginal states denoted as $\rho_A$ and $\rho_B$, we can introduce the symmetric linear mutual information
\begin{equation}
I_L (\rho_{AB}) := \parallel \rho_{AB} - \rho_{A}\otimes \rho_{B}\parallel _2 ^2 = tr((\rho_{AB} - \rho_{A}\otimes \rho_{B})^2)
\end{equation}
as a measure of the total correlation in the state $\rho_{AB}$.
$I_L (\rho_{AB})$ is always non-negative, it vanishes only when $\rho_{AB}$ is a product state and reaches the maximum $1-\frac{1}{d^2}$ for a fixed composite Hilbert space when $\rho_{AB}$ is a maximally entangled state.

In order to reveal the correlation of $\rho_{AB}$ in a particular basis $\{ \left| i_{\theta} \right\rangle\}$ of system A, we  project Alice's system onto this basis  and denote the overall state after the local measurement as $\rho_{\theta B}$ (which is given in (\ref{totalstateafterlocalmeasontheta})).
We use $I_L (\rho_{\theta B})$, the linear mutual information of the resulting state $\rho_{\theta B}$, as a measure of the correlation of $\rho_{AB}$ with respect to the local basis $\{ \left| i_{\theta} \right\rangle\}$ of system A.

The correlation of the state $\rho_{AB}$ on a $d\otimes d$  Hilbert space can be revealed in different bases. The sum of the correlations revealed in different MUBs is conserved, and we have the following theorem.

{\bf Theorem 2.} When a full set of $d+1$ local MUBs exists, the sum of the correlations revealed in a full set of MUBs is equal to the total correlation in the bipartite state, i.e.,
\begin{equation}
\sum_{\theta=1}^{d+1} I_L(\rho_{\theta B}) = I_L (\rho_{AB}) .
\end{equation}

The proof is left to the Supplementary Information.

\end{document}